\documentclass[aps,prb,twocolumn,superscriptaddress]{revtex4-1}

\usepackage{graphicx}
\usepackage{amsmath}
\usepackage{color}

\newcommand{\VSD}{\ensuremath{V_{\text{SD}}}}

\newcommand{\TMC}{\ensuremath{\mathcal{T}}}

\newcommand{\tc}{\ensuremath{t_\mathrm{c}} }

\newcommand{\Tstar}{\ensuremath{T_2^*} }
\newcommand{\kB}{\ensuremath{k_\mathrm{B}} }

\begin{document}
\title{Photon-assisted tunneling and charge dephasing in a carbon nanotube double quantum dot}

\author{A. Mavalankar}
\affiliation{Department of Materials, University of Oxford, Parks Road, Oxford OX1 3PH, United Kingdom}
\author{T.Pei}
\affiliation{Department of Materials, University of Oxford, Parks Road, Oxford OX1 3PH, United Kingdom}
\author{E. M. Gauger}
\affiliation{SUPA, Institute of Photonics and Quantum Sciences, Heriot-Watt University, EH14 4AS, United Kingdom}
\author{J. H. Warner}
\affiliation{Department of Materials, University of Oxford, Parks Road, Oxford OX1 3PH, United Kingdom}
\author{G.A.D. Briggs}
\affiliation{Department of Materials, University of Oxford, Parks Road, Oxford OX1 3PH, United Kingdom}
\author{E.A. Laird}
\affiliation{Department of Materials, University of Oxford, Parks Road, Oxford OX1 3PH, United Kingdom}

\begin{abstract}

We report microwave-driven photon-assisted tunneling in a suspended carbon nanotube double quantum dot.  From the resonant linewidth at a temperature of 13~mK, the charge dephasing time is determined to be $280\pm30$~ps. The linewidth is independent of driving frequency, but increases with increasing temperature. The moderate temperature dependence is inconsistent with expectations from electron-phonon coupling alone, but consistent with charge noise arising in the device. The extracted level of charge noise is comparable with that expected from previous measurements of a valley-spin qubit, where it was hypothesized to be the main cause of qubit decoherence. Our results suggest a possible route towards improved valley-spin qubits.

\end{abstract}

\date{\today}

\maketitle

\section{Introduction}

Qubit manipulation in carbon nanotube devices is of interest both for quantum computing~\cite{Churchill2009,Flensberg2010, Palyi2011, Pei2012} and for coherent control of nanomechanical states via strong coupling to a spin qubit~\cite{Palyi2012, Ohm2012}.  The low concentration of nuclear spins in carbon is at first sight promising for spin and valley coherence, but the qubit coherence time observed so far~\cite{Laird2013, Viennot2014} has not exceeded 100~ns, well below the expected limit from electron-lattice and spin-orbit coupling~\cite{Bulaev2008, Rudner2010}. Instead, it is possible that this coherence is limited by charge dephasing, with a coupling to valley-spin states arising from spin-orbit coupling or gate-voltage dependent inter-dot exchange~\cite{Li2014, Laird2015}. 

Here, we probe charge dephasing by measuring photon-assisted tunneling (PAT)~\cite{Meyer2007} for the first time in a carbon nanotube double quantum dot.  This effect occurs when energy absorbed from a microwave electric field stimulates charge tunneling between quantum dots that would otherwise be suppressed by Coulomb blockade. Because PAT is a resonant effect, the measured width of the absorption line is proportional to the charge dephasing rate.
By measuring the frequency and temperature dependence of the PAT resonance linewidth, we deduce that in this device, low-temperature charge dephasing is limited mainly by electrical noise, with possible contribution from phonon coupling above 50~mK, and with electron tunneling excluded as the dominant mechanism. The noise level deduced from the dephasing rate is of the same order of magnitude as estimated previously to account for dephasing of the nanotube valley-spin qubit~\cite{Laird2013}.

\section{Photon-assisted tunneling}

\begin{figure}
\includegraphics[width=\columnwidth]{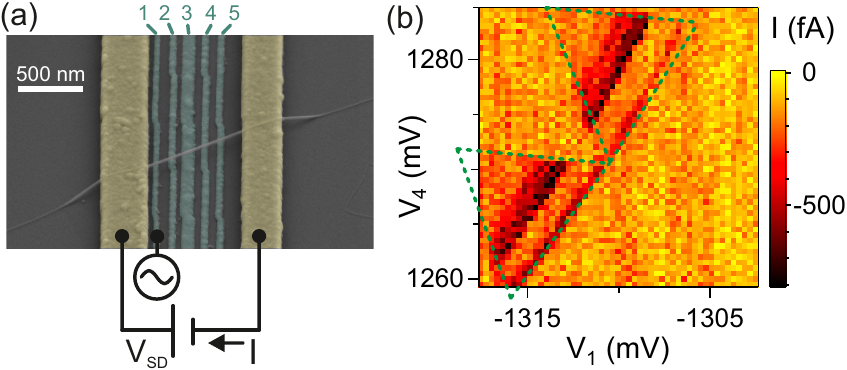}
\caption{\label{Fig1} (Color online) A carbon nanotube double quantum dot (a) Scanning electron microscope image of a device lithographically similar to the one measured. A carbon nanotube is stamped across two contact electrodes, spanning a trench in which five gates (labelled 1-5) are defined. Through the combination of Schottky barriers near the contacts and voltages applied to the gates, a double quantum dot potential is formed. A source-drain bias $\VSD$ applied to the contact electrodes leads to a measured current~$I$. Photon-assisted tunneling can be driven by applying a microwave voltage to gate 1. For the image, $\sim 2$~nm Pt was evaporated to improve nanotube visibility. (b) Current through the device measured as a function of gate voltage near a double quantum dot transition with $\VSD=-1$~mV and no microwaves applied. The two triangles of allowed current are indicated.}
\end{figure}

To fabricate the double quantum dot [Fig.~\ref{Fig1}(a)], a carbon nanotube is synthesized by chemical vapour deposition on a quartz chip using FeCl as a catalyst~\cite{Zhou2008}. On a separate chip, consisting of intrinsic Si substrate with 300~nm thermal oxide, 20~nm thick Cr/Au gates (labelled 1-5) are patterned, followed by 130~nm thick Cr/Au contact electrodes to define a trench. The nanotube is transferred by stamping~\cite{Wu2010, Pei2012} so that it is suspended over the trench between the two contacts. The device is measured in a dilution refrigerator with a base temperature of 13~mK. Electron thermalization is achieved by heat-sinking all dc via printed circuit board copper powder filters containing $RC$ stages with $\sim100$~kHz cutoff~\cite{Mueller2013}. To allow a microwave signal to be added to the DC gate voltage, Gate~1 is connected via a bias tee to a microwave source via a section of NbTi superconducting coaxial cable fitted with cold attenuators totalling 28~dB.

A double quantum dot defined through a combination of Schottky barriers at the contacts and DC gate voltages, which together deplete three short segments of the nanotube so as to create a double-well potential. The gate voltages are set so that the double dot is measured in the $pp$ configuration, where both quantum dots are occupied by holes. The chemical potentials in the left and right dots are adjusted by sweeping gate voltages $V_1$ and $V_4$, applied to gates 1 and 4 respectively. Hole tunneling through the three tunnel barriers gives rise to a currnet $I$ through the nanotube. With a bias $\VSD$ applied between the contacts, and with microwaves turned off, the current is shown in Fig.~\ref{Fig1}(b) as a function of $V_1$ and $V_4$.  For most gate voltage settings, the alignment of energy levels in the two quantum dots means that hole tunneling is suppressed by Coulomb blockade. The two triangles of allowed current are characteristic of a double quantum dot~\cite{Wiel2003}.

\begin{figure}
\includegraphics[width=\columnwidth]{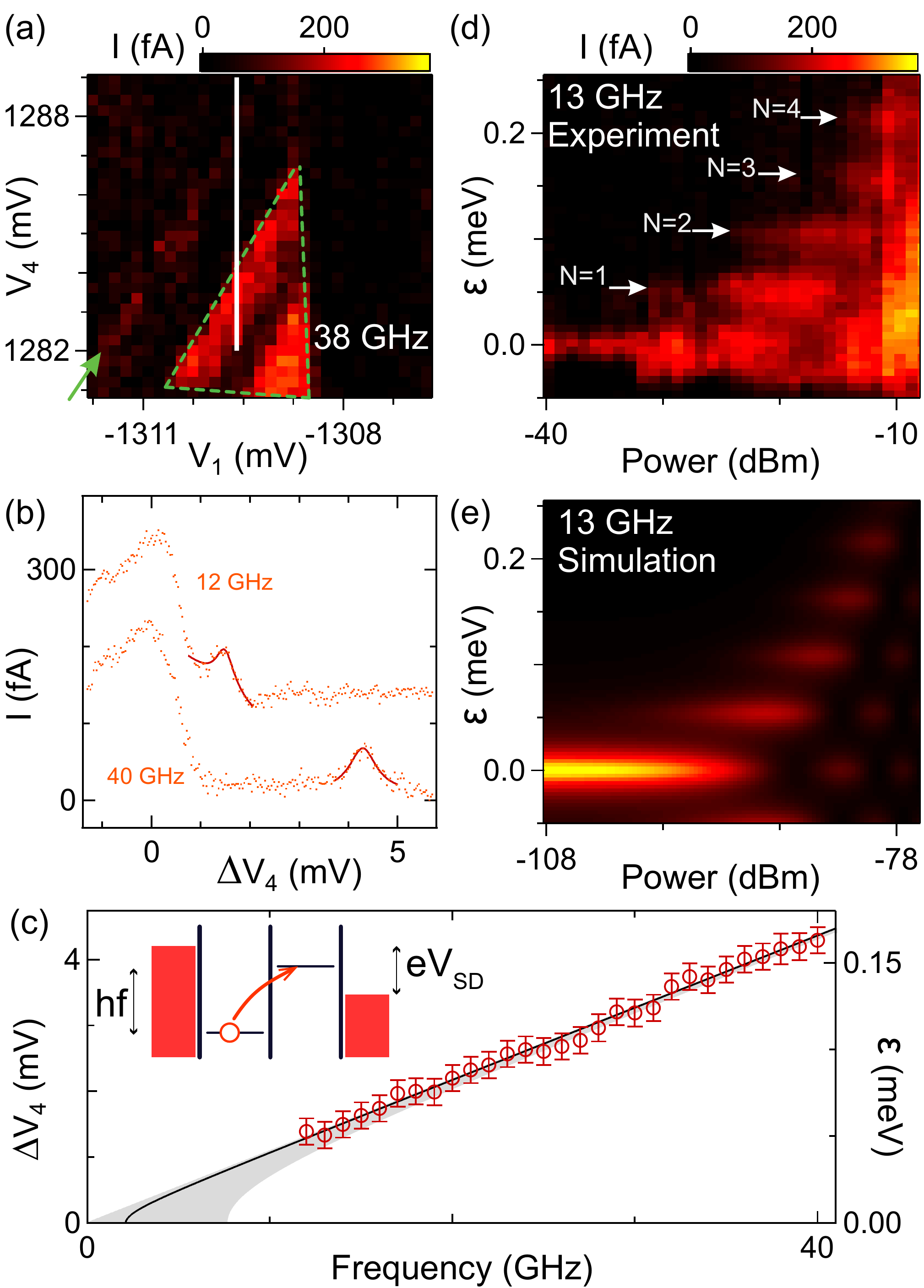}
\caption{\label{Fig2} (Color online) Photon-assisted tunneling (a) Current as a function of gate voltages with bias $\VSD=0.2$~mV and microwaves at 38 GHz. A PAT peak (marked with arrow) is evident as a satellite line outside the main transport triangle. The detuning axis used below is marked by a vertical line. (b) Points: current as a function of detuning for different microwave frequencies. Lines: Fits to the PAT peaks. The 12~GHz data is offset vertically for clarity. (c) PAT spectroscopy. Symbols: peak detuning as a function of frequency, extracted as in~(b). Error bars reflect uncertainty in the gate voltage corresponding to zero detuning as well as in extracted fit parameters. Line: Fit to two-level model, yielding interdot tunneling $\tc=4\pm7~\mu eV$. The $\pm1\sigma$ confidence band is shaded. Right axis marks the detuning $\varepsilon$ corresponding to $\Delta V_4$, related by the lever arm $\lambda$ extracted from the data. Inset: hole chemical potential of dots and leads on resonance, showing the Coulomb-blocked PAT transition. (d) Current as a function of detuning and source microwave power at 13~GHz. The series of PAT fringes is marked. (e) Simulated current with $\Tstar=300$~ps and assuming perfect coupling of the microwave voltage to the double dot detuning.}
\end{figure}

To detect PAT, the current is measured with microwaves applied [Fig.~\ref{Fig2}(a)]. A line of microwave-induced current [marked by an arrow in Fig.~\ref{Fig2}(a)] appears outside the transport triangle, in a region of gate space where Coulomb blockade suppresses current in the absence of microwaves. For these gate voltage settings, Coulomb blockade is broken by photon absorption~\cite{Oosterkamp1998} allowing a hole to tunnel uphill from left to right between two resonant levels and subsequently escape to the right lead [Fig.~\ref{Fig2}(c) inset]. Measuring the position of this line as a function of microwave frequency $f$ allows spectroscopy of the double quantum dot. With the detuning $\varepsilon$ defined as the difference in hole chemical potentials between right and left dots, a current peak is expected whenever $Nhf = \Delta E (\varepsilon)$, where $\Delta E(\varepsilon)$ is the energy difference between bonding and antibonding states of the double dot, $N$ is an integer, and $h$ is Planck's constant. 

This PAT spectroscopy is seen in Fig.~\ref{Fig2}(b), which shows $N=1$ peaks for two different microwave frequencies. The current is plotted against gate voltage location $\Delta V_4$, defined along the line marked in Fig.~\ref{Fig2}(a). With the origin of this $\Delta V_4$ set at the triangle baseline, where the chemical potentials of left and right dots are equal, detuning is related to gate voltage by $\varepsilon = \lambda \Delta V_4$, where $\lambda$ is the lever arm between gete voltage and detuning. As expected, the peak moves towards greater detuning with higher frequency. To characterize this behaviour in more detail, we fit the current peak at each frequency with a Lorentzian, including an offset slope to account for background current. The fitted peak centers are plotted in Fig.~\ref{Fig2}(c). The dependence of peak detuning on frequency is approximately linear, consistent with a weakly tunnel-coupled double dot. The data are fitted to the equation expected from a two-level model~\cite{Oosterkamp1998}:
\begin{equation}
\Delta V_4 = (\lambda e)^{-1} \sqrt {(hf)^2 - 4\tc^2},
\end{equation}
where $e$ is the electron charge, with $\lambda$ and the inter-dot tunnel coupling $\tc$ taken as fit parameters. As seen, the spectroscopic effect of the tunnel coupling $\tc=4\pm7~\mu$eV is negligible within the uncertainty. The lever arm $\lambda = 0.038\pm0.002$ is consistent with that obtained from the size of the bias triangle in Fig.~\ref{Fig2}(a)~\cite{Wiel2003}.

With increased microwave power, multi-photon transitions can be observed [Fig.~\ref{Fig2}(d)]. The dependence on detuning and power is in reasonable agreement with a model of resonant transitions with dephasing~\cite{Stoof1996} [Fig.~\ref{Fig2}(e) and Appendix~\ref{ap:multiphoton}]. In particular, the dependence of current on power is weakly non-monotonic, an indication of interference between different coherent transitions. However, this effect is masked by a non-resonant current increase at high power, presumably reflecting incoherent tunneling due to heating. The power offset between simulation and data allows the ratio between the generator output and quantum dot detuning to be extracted, which at this frequency is 68~dB. For the following data the power is set well below the threshold for multi-photon processes.

\section{Charge dephasing}

\begin{figure}
\includegraphics[width=\columnwidth]{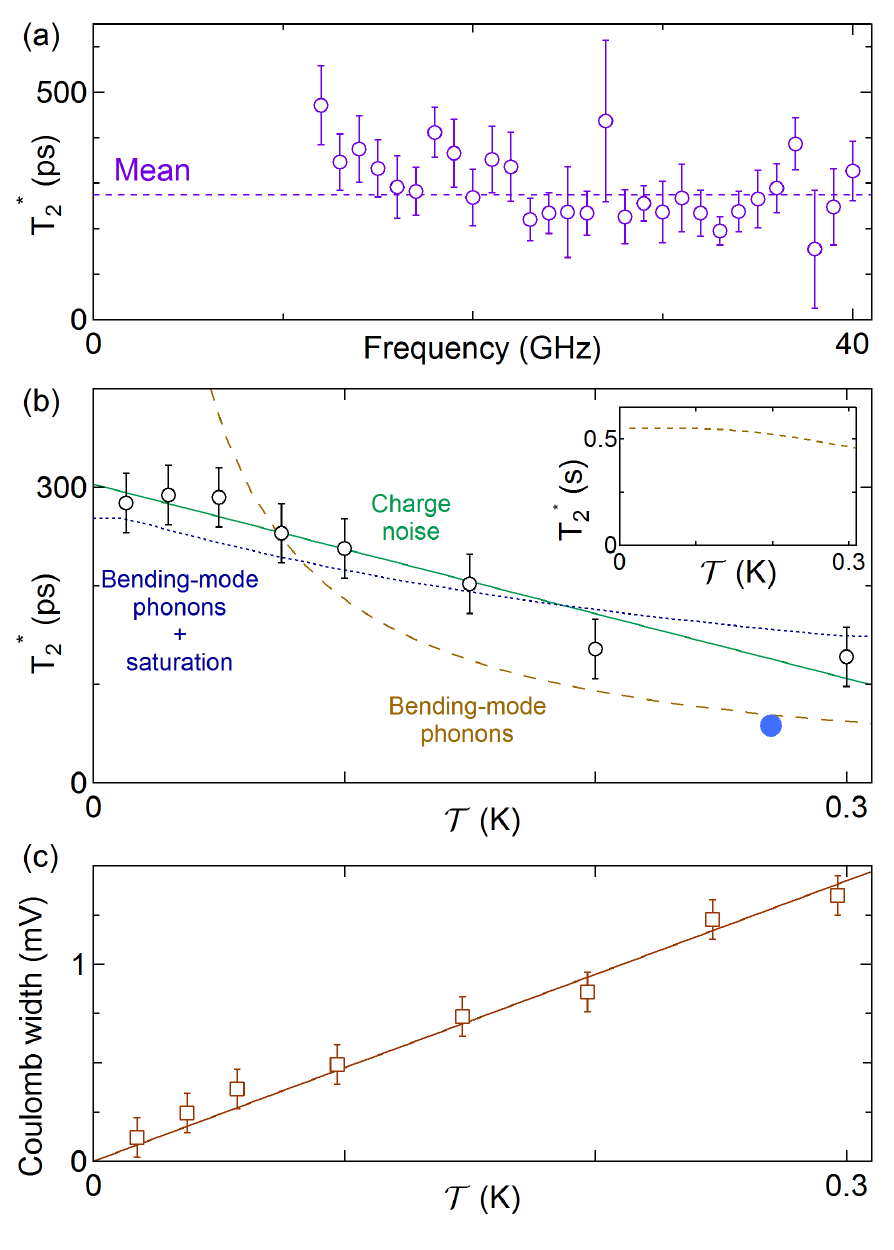}
\caption{\label{Fig3} (Color online) (a) Charge dephasing time $\Tstar$, extracted from fitted PAT linewidths, as a function of frequency at 13~mK. (b) Open symbols: Average $\Tstar$ as a function of fridge temperature $\TMC$. Dashed curve: Fit to a model of dephasing (see text) from bending-mode phonons. The two points at lowest $\TMC$ are excluded because they provide a very poor match to the model. Dotted curve: Fit to phonon model combined with saturation. Solid curve: Fit to a model of thermally activated charge noise. Filled symbol: Expected charge $\Tstar$ arising from estimated charge noise in the valley-spin qubit experiment of Ref.~\onlinecite{Laird2013}. Inset: predicted dephasing by stretching modes (see text).  (c) Measured Coulomb step width (symbols) as a function of fridge temperature. The data are well fit by a straight line through the origin, confirming electron thermalization down to at least 20~mK.}
\end{figure}

By studying PAT as a function of frequency and fridge temperature $\TMC$, information about charge dephasing and hence about the electrical environment can be obtained. The dephasing time $T_2^*$ is extracted from the width of the PAT peak in detuning~\cite{Petta2004} using the formula $\Tstar = 2h/\lambda e \gamma_V$, where $\gamma_V$ is the full width at half maximum of the Lorentzian. Figure \ref{Fig3}(a) shows $\Tstar$ as a function of microwave frequency at $\TMC=13$~mK, obtained from fits as in Fig.~\ref{Fig2}(b). The mean value is $\Tstar= 280\pm 30$~ps, similar to measurements in GaAs quantum dots~\cite{Hayashi2003, Petta2004, Frey2012}, Si donors~\cite{Dupont-Ferrier2013}, and nanotube circuit-quantum-electrodynamics devices~\cite{Viennot2015, Ranjan2015}. Both in this data and at higher temperatures (not shown), no strong dependence on frequency is seen. Figure \ref{Fig3}(b) shows the temperature dependence of $\Tstar$ averaged over the entire measured frequency range. As expected, $\Tstar$ decreases at higher temperature, indicating thermally activated dephasing mechanisms.

To confirm proper thermalization of the device, Fig.~\ref{Fig3}(c) shows the measured width of a Coulomb step feature (the edge of a bias triangle similar to Fig.~\ref{Fig2}(a)) as a function of temperature with no microwaves applied. Because this step is broadened by thermal smearing in the leads, the linear dependence indicates that the electron temperature closely tracks $\TMC$. 

\section{Mechanisms of dephasing}

In this section, we compare the data of Fig.~\ref{Fig3}(a-b) with expectations from various charge dephasing processes in this device~\cite{Hayashi2003}.
We consider first vibrational dephasing, by analysing the phonon spectrum and dominant coupling mechanisms to the charge state. 
We must consider both bending and stretching modes as potential contributors to charge dephasing. For the former, coupling to the charge state arises from displacement of the quantum dots in the gate electric field, whereas the electron-phonon coupling of longitudinal phonons arises from the deformation potential. The spectra of bending and stretching phonons are (neglecting tension) given by~\cite{Suzuura2002, Ando2005, Bulaev2008, Rudner2010}:
\begin{align}
\omega_{B_n}  &= \pi^2 c_S R / \sqrt{2} L^2\times n^2~,  \label{eq:wnB} \\
\omega_{S_n} & = \pi c_S / L \times n~,\label{eq:wnS}
\end{align}
where $c_S \approx 2 \times 10^4~\mathrm{ms}^{-1}$ is the longitudinal sound velocity and $R$ and $L$ are the radius and length of the suspended nanotube. Estimating $R=2$~nm and using $L=650$~nm results in $\omega_{B_1}/2\pi = 105$~MHz and $\omega_{S_1}/2\pi = 15.4$~GHz for the lowest frequency bending and stretching mode. 

The respective electron-phonon couplings for bending and stretching modes are calculated in Appendix~\ref{ap:SD}:
\begin{align}
g_B(\omega_n) &= \kappa_{B} \omega_n^{-1}~, \label{eq:gB} \\
g_S(\omega_n) &= \kappa_{S} \omega_n^{-1/2} ~.
\label{eq:gS}
\end{align}
Numerical estimates considering the underlying mechanisms~\cite{Supp} give $\kappa_B \sim 5\times 10^{19}~\mathrm{s}^{-2}$ (assuming an electric field $E\sim 10^7~\mathrm{Vm}^{-1}$ based on applied gate voltage and sample geometry, and nanotube mass $m\sim 6 \times 10^{-21}$~kg), and $|\kappa_S| \sim2\times 10^{13}~\mathrm{s}^{-3/2}$. In both cases the coupling strength peaks for the lowest frequency mode.

Dephasing will thus be dominated by thermally populated low-frequency modes. Contrary to a previous treatment of double dot electron-phonon coupling~\cite{Hayashi2003} based on the canonical spin-boson model with quasi-continuous spectral density~\cite{Leggett1987}, we anticipate that only a small number of modes is relevant in this case, suggesting  a different approach is required. To estimate $\Tstar$, we therefore adapt analytically derived Rabi-model dephasing rates~\cite{Fruchtman2015}. 
These rely upon a perturbative expansion utilising the phase space representation of the oscillators. In our case the charge dephasing rate due to a single phonon mode $\omega$ with coupling $g$ is
\begin{equation}
\begin{split}
(\Tstar)^{-1} = \frac{2 g^2 \gamma}{\hbar^2\Omega^2}
\left(
   \frac{\varepsilon^2}{\frac{\gamma^2}{4}+\omega^2}
+ \frac{t_c^2}{\frac{\gamma^2}{4} + (\Omega - \omega)^2}
\right.
\\
\left.
+ \frac{t_c^2}{\frac{\gamma^2}{4} + (\Omega +\omega)^2}
\right)
\coth \frac{\hbar\omega}{2\kB \TMC}
\label{eq:Q}
\end{split}
\end{equation}
with $\kB$ being Boltzmann's constant, $\gamma$ the phonon relaxation rate, and $\hbar\Omega = \Delta E(\varepsilon) = \sqrt{\varepsilon^2 + 4t_c^2}$ the charge qubit energy splitting. In contrast to the spin-boson model, where the loss of coherence is determined by the spectral density~\cite{Brandes2005}, Eq.~(\ref{eq:Q}) explicitly includes the finite lifetime of the relevant phonon modes, as well as the charge bias and tunneling. Generalisation to multiple modes is straightforward~\cite{Fruchtman2015} but unnecessary here: for the parameters discussed below inclusion of up to 10 modes gives relative changes of less than $10^{-4}$ (Appendix~\ref{ap:approximation}).

Figure~3(b) shows a fit of $\Tstar$ calculated using~Eq.~(\ref{eq:Q}) for the lowest frequency bending mode, where $\kappa_B$ in Eq.~(\ref{eq:gB}) is taken as the fitting parameter. For the frequency of the lowest mode, measured electromechanically~\cite{Sazonova2004,Ares2015}, we take $\omega=2\pi \times 300$~MHz and from the linewidth $\gamma^{-1} = 500~\mathrm{ns}$. The higher value of $\omega$ than in Eq.~(\ref{eq:wnB}) presumably reflects tension in the suspended nanotube~\cite{Sapmaz2003}. The fit yields $\kappa_B\approx 3.5 \times 10^{19}~\mathrm{s}^{-2}$, in reasonable agreement with the previous estimate given the many uncertainties. By contrast, the lowest stretching mode [using Eqs.~(\ref{eq:wnS}) and (\ref{eq:gS})] leads to dephasing on a timescale approaching seconds [Fig.~3(b)inset] and can therefore be ruled out. 
The calculated $\Tstar$ shows almost no dependence on $\Omega$ in the measured temperature range, consistent with  Fig.~3(a).
Bearing in mind that Eq.~(\ref{eq:Q}) relies on a perturbative expansion of electron-phonon coupling, we have checked that full numerical simulations of a single mode Rabi model using QuTip~\cite{QuTip} lead to a temperature dependence in the dephasing time that is consistent with Eq.~(\ref{eq:Q})~(Appendix~\ref{ap:approximation}).

Thus, while bending mode phonons may limit $\Tstar$ above $\TMC\approx 50$~mK, they cannot explain the measurements across the entire temperature range due to the strong temperature dependence of the phonon model. This suggests that at least at low temperatures a different mechanism limiting coherence must be at play. We therefore next compare with expectations from electrical noise. In this case, the dephasing time is given by $\Tstar = \sqrt{2} h / \lambda \Delta \varepsilon_\mathrm{rms}$, where $\Delta \varepsilon_\mathrm{rms}$ is the root-mean-square detuning noise at frequencies up to the driving frequency~\cite{Laird2013}. Assuming that the noise spectrum is dominated by frequencies below 10~GHz, this is consistent with the data in Fig.~3(a) if $\Delta \varepsilon_\mathrm{rms} = 0.02$~meV at low~$\TMC$. There is no clear expectation for the temperature dependence; however, in a GaAs spin qubit device where dephasing was attributed to thermally activated electrostatic noise, an approximately linear dependence of $\Tstar$ on $\TMC$ was found~\cite{Dial2013}. This simple model fits the data well [Fig.~3(b)], meaning charge noise cannot be excluded as the limit of $\Tstar$ across the temperature range.

Thirdly, we suppose that phonon-mediated dephasing operates in conjunction with some other mechanism, such as temperature-independent charge noise. To model these two mechanisms combining incoherently, we fit the data in Fig.~\ref{Fig3}(b) across the entire range with
\begin{equation}
\Tstar = (1/T_{2,\mathrm{sat}}^* + 1/T_{2,\mathrm{p}}^*)^{-1},
\end{equation}
where $T_{2,\mathrm{p}}^*$ is the phonon dephasing time given by Eq.~(\ref{eq:Q}) and $T_{2,\mathrm{sat}}^*= 280$~ps is a saturation value taken as equal to the measured $\Tstar$ at 13~mK. This model also fits the data well with $\kappa_B = 1.5 \times 10^{19}~\mathrm{s}^{-2}$.

Finally, we consider that $\Tstar$ is set by the lifetime of the excited state, limited by tunneling to the leads. Because escape from the right dot is not Coulomb-blocked [Fig.~2(c) inset], this effect should show no strong temperature dependence, and therefore is not the main limit on~$\Tstar$. Furthermore, we observe no dependence of $\Tstar$ on gate voltage tuning, indicating that it is not affected by tunnel barriers as tunneling or cotunneling would be. In conclusion, we deduce that charge coherence is limited by electrical noise at low temperatures, with a possible contribution from phonons above 50~mK.

\section{Conclusion}

Because PAT spectroscopy gives a direct measurement of the charge dephasing rate, it may shed light on the cause of dephasing for a valley-spin qubit defined in a similar device~\cite{Laird2013}. In that experiment, performed at 270~mK, a voltage-dependent spin splitting made the qubit sensitive to electrical noise, which was suggested as the main limit on the decoherence rate. To explain the measured valley-spin decoherence rate, an rms detuning jitter $\varepsilon_\mathrm{rms} \sim 0.1$~meV was required. In Fig.~\ref{Fig3}(b), the charge~$\Tstar$ expected from the same detuning jitter is plotted as a filled circle, and is found to be of the same order of magnitude as measured here. It is therefore plausible that charge noise was a contributor to the decoherence and dephasing rates measured for that valley-spin qubit~\cite{Laird2013}. 

The origin of the noise remains unclear. The required noise level is far in excess of expectation from our room-temperature electronics. The fact that the noise is apparently reduced compared with that in Ref.~\cite{Laird2013} (where the nanotube rested on an oxide layer) suggests a contribution from the substrate. However, in our device most substrate noise should be screened by the surface gates. One origin could be electrostatic patch noise on the electrode surfaces, as seen in ion traps~\cite{Wineland1998}. Another is possible presence of fluctuating charge traps on the surface of the nanotube itself, either in amorphous carbon deposited during synthesis, or from adsorbed water~\cite{Kim2003}. These results suggest that, provided other decoherence mechanisms such as magnetic~$^{13}$C nuclear impurities can be eliminated, careful control of nanotube synthesis and/or vacuum conditions in the cryostat may enhance both charge and valley-spin coherence in nanotube quantum dots.

\section{Acknowledgements}

We acknowledge N.~Ares, C.S.~Allen, and Y.~Li for discussions, and support from EPSRC (EP/J015067/1), Templeton World Charity Foundation, a Marie Curie Career Integration Grant, the Royal Society of Edinburgh and Scottish Government, and the Royal Academy of Engineering.

\appendix

\section{Photon-assisted tunneling with dephasing}
\label{ap:multiphoton}
The simulation in Fig.~2(e) uses a model of time-dependent resonant tunneling between two states \cite{Stoof1996}, with the phenomenogical inclusion of a level-broadening term due to dephasing. The current is:
\begin{equation}
I(V_\mathrm{ac}, \varepsilon) 
= A \sum_n
\frac{J_n^2(\varepsilon_\mathrm{ac}/2\pi h f)}
{(2\pi n h f - \varepsilon^2) + (2h/\Tstar)^{2}},
\end{equation}
where $V_\mathrm{ac}$ is the microwave voltage at the signal generator, $\varepsilon_\mathrm{ac}=\alpha e V_\mathrm{ac}$ is the corresponding detuning variation, and $A$ an overall scaling factor (set by the double-dot tunnel couplings). Here $\alpha$ is the insertion loss between generator and device, including attenuators, cable losses, and the gate lever arm. To generate Fig.~2(e) $\Tstar$ was set to 300~ps and $\alpha$ and $A$ were adjusted by hand to match the data. By this procedure $\alpha$ is estimated at 68~dB, consisting of 28~dB from the inline attenuators, $-20 \log \lambda \sim 28$~dB from the lever arm, and $\sim 12$~dB from losses in room-temperature cable and the sample holder.

\section{Electron-phonon coupling constants}
\label{ap:SD}

Our starting point is the spin-boson Hamiltonian~\cite{Leggett1987, Brandes2005}:
\begin{equation}
H_\mathrm{SB} = H_\mathrm{e} + H_\mathrm{ph} + H_\mathrm{e-ph} ~, \label{eq:HSB}
\end{equation}
with bare double dot represented by a pseudo-spin Hamiltonian 
\begin{equation}
H_\mathrm{e} = \frac{\varepsilon}{2}\sigma_z + t_\mathrm{c} \sigma_x ~,
\end{equation}
where $\sigma_z = \vert L \rangle \langle L \vert - \vert R \rangle \langle R \vert$ and $\sigma_x = \vert L \rangle \langle R \vert + \vert R \rangle \langle L \vert$ for $\vert L / R \rangle$ representing the presence of the hole on the left or the right dot, respectively. Each phonon mode with wavevector $q$, angular frequency $\omega_q$, and ladder operators $a_q, a^\dagger_q$ is governed by the Hamiltonian of the harmonic oscillator   
\begin{equation}
H_\mathrm{ph}=\sum_q \hbar \omega_q a_q^\dagger a_q ~,
\end{equation}
and the charge-phonon coupling given by
\begin{equation}
H_\mathrm{e-ph} = \sigma_z \sum_q \hbar g(\omega_q) \left( a_{q} + a^\dagger_{q} \right ).
\label{Heph}
\end{equation}
Here the electron-phonon coupling in each mode is parameterized by $g(\omega_q)$.

To estimate $g(\omega)$ we express the oscillator displacement for each wavevector $q$ in terms of the (real-space) position variable $u_q$, 
\begin{equation}
a_{q} + a^\dagger_{q}= \sqrt{\frac{2m\omega_q}{\hbar}}u_q ~,
\end{equation} 
where $m$ is the mass of the nanotube. Two physical mechanisms contribute to the electron-phonon interaction: coupling to bending-mode phonons via the gate electric field and coupling to longitudinal phonons via the deformation potential~\cite{Ando2005,Suzuura2002,Rudner2010,Bulaev2008}. We estimate them in turn.

\subsection{Bending-mode phonons}
The bending-mode dispersion relation~\cite{Bulaev2008} is $\omega_q = c_SRq^2/\sqrt{2}$, with $q=n\pi/L$, where $R$ and $L$ are the radius and length of the nanotube. The deformation potential associated with these phonons averages to zero around the nanotube circumference~\cite{Bulaev2008}, so   electron-phonon coupling arises mainly from the perpendicular electric field $E$ induced by the gates.
This mediates coupling to bending-mode phonons according to the Hamiltonian
\begin{equation}
H^B_\mathrm{e-ph} = \sum_q eE F(q) u_q \sigma_z,
\label{HephE}
\end{equation}
where $F(q)\sim 2/qL$ is a form factor that parameterizes the average displacement over the length of a quantum dot~\cite{Golovach2004}. Equating Eq.~(\ref{HephE}) with Eq.~(\ref{Heph}) gives
\begin{equation}
g_B(\omega)\sim\frac{2^{5/4}eE}{\omega L}\sqrt{\frac{c_S R}{\hbar m}}.
\end{equation}

\subsection{Longitudinal phonons}
For longitudinal phonons, we must consider both stretching and torsional modes, with each mode coupling via both diagonal and off-diagonal matrix elements in sublattice space~\cite{Ando2005}. The dominant term is the diagonal coupling to stretching-mode phonons, whose dispersion relation is $\omega_q=c_S q$. The deformation potential Hamiltonian in this case leads to~\cite{Bulaev2008}
\begin{equation}
H^S_\mathrm{e-ph} = \sum_q iG q F(q) u_q \sigma_z.
\label{HephS}
\end{equation}
where the coupling constant is $G \sim 21$~eV. Equating Eqs.~(\ref{HephS}) and (\ref{Heph}) gives
\begin{equation}
g_S(\omega)\sim i\frac{2^{3/2}G}{L}\sqrt{\frac{1}{m \omega \hbar}}.
\end{equation}
For torsional modes, the dispersion relation is similar and the coupling constant is about an order of magnitude smaller~\cite{Bulaev2008}. These can therefore be neglected. 

Since $g_B$ and $g_S$ are out of phase, the mechanisms decouple to second order and can be treated independently.

\section{Approximation of dephasing time}
\label{ap:approximation}

\begin{figure}
\includegraphics[width=\columnwidth]{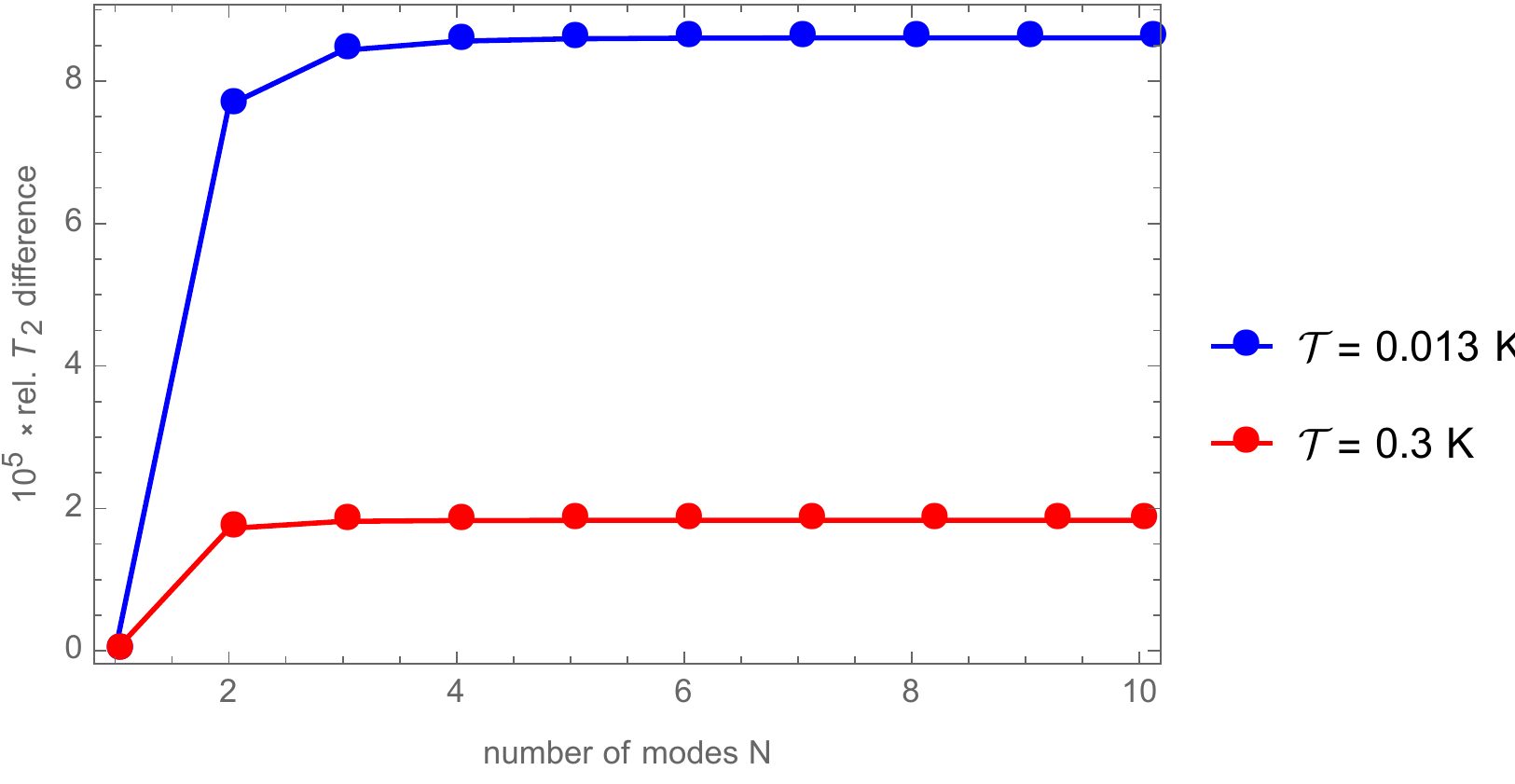}
\caption{\label{fig:relerr} (Color online) Relative shortening of the dephasing time as more than one modes are considered.}
\end{figure}

Ref.~\onlinecite{Fruchtman2015} allows us obtain an analytical estimate of the `spin' coherence time in the presence of one or more oscillator modes. This is more appropriate than solutions to the canonical spin-boson model, as the discrete oscillator environment is too small to assume short environmental correlations times. By contrast, the phase space representation technique~\cite{Fruchtman2015}  is capable of resolving non-Markovian dynamics of combined charge qubit and oscillator(s). Note that in our case the qubit dephasing will be predominantly caused by oscillator relaxation.

The main text only considers the lowest frequency mode. To verify the validity of this simplification, Fig.~\ref{fig:relerr} shows the effect on the predicted dephasing time as the next higher frequency modes are included at both the lowest and the highest temperature of the experiment. In both cases the relative shortening of the dephasing time is smaller than $10^{-4}$.

At low temperature, we can also solve the full dynamics of Hamiltonian~(\ref{eq:HSB}) numerically, for example using a package like QuTip~\cite{QuTip}. However, rigorously extracting a dephasing time is not straightforward: time traces of the charge qubit's coherence are rich in features over a long time in the relevant parameter regime, leading to a rather messy Fourier spectrum. A possible way of extracting a crude estimate of the dephasing time is to consider the amplitude range of coherence oscillations over ten Rabi periods $\xi(\tau) = \max | \rho_{01}(t)|~\mathrm{for}~t \in [\tau,  \tau + 10 \times 2 \pi / \Omega]$,  and finding the smallest $\tau$ for which $\xi(\tau) / \xi(0) < 1/ e$. 

In the low temperature regime $\TMC < 0.1$~K (where the Hilbert space can be safely truncated at below 100 excitations) a single mode Rabi model with parameters as in the main text shows a temperature dependence in the decay of the qubit coherence that is consistent with the predictions of Eq.~(6) in the main text. 
We therefore conclude that the qualitative predictions of Eq.~(6) are indeed adequate for ruling out phonons as the origin of low temperature charge dephasing.

\vfill


%

\end{document}